\documentclass[apj]{emulateapj}

\def\ergs{erg~s$^{-1}$}
\def\ergcms{erg~cm$^{-2}$~s$^{-1}$}

\begin{document}

\title{The Optical Counterpart of NGC 1313 X-1}

\author{Lin Yang\altaffilmark{1,2}, Hua Feng\altaffilmark{3}, and Philip Kaaret\altaffilmark{4}}

\altaffiltext{1}{Undergraduate Student Research Training Program, Department of Engineering Physics, Tsinghua University, Beijing 100084, China}
\altaffiltext{2}{Physics Department and Center for Astrophysics, Tsinghua University, Beijing 100084, China}
\altaffiltext{3}{Department of Engineering Physics and Center for Astrophysics, Tsinghua University, Beijing 100084, China}
\altaffiltext{4}{Department of Physics and Astronomy, University of Iowa, Van Allen Hall, Iowa City, IA 52242, USA}

\shorttitle{The Optical Counterpart of NGC 1313 X-1}
\shortauthors{Yang, Feng, \& Kaaret}

\begin{abstract}
We identify the optical counterpart of the ultraluminous X-ray source (ULX) NGC 1313 X-1 and discuss constraints on its physical nature from multiband optical spectra.  There is a single object on Hubble Space Telescope (HST) images within the aspect-corrected Chandra X-ray error circle; a fainter, possibly extended, feature lies near the edge of the error circle.  The brighter object showed prominent variation in the F555W band, but was constant in the F814W band. The spectrum was consistent with a single power-law on 2003 Nov 17, but deviated from this on 2004 Jul 17, suggestive of more than one emission component.  Based on the location, magnitudes, spectral shape, and variability of the bright object, it is likely the ULX counterpart.  The red wing of the spectrum around F814W may be due to emission from the companion star, and the blue wing is likely from disk emission. The stellar population around X-1 has an age older than 30~Myr, without very blue stars or young clusters. This places a constraint on the companion mass of the ULX as no more than 10~$M_\sun$.
\end{abstract}

\keywords{black hole physics --- accretion, accretion disks --- X-rays: binaries --- X-rays:Individual (NGC 1313 X-1)}

\section{Introduction}

Ultraluminous X-ray sources (ULXs) are variable, nonnuclear X-ray sources found in nearby galaxies with isotropic luminosities higher than $3 \times 10^{39}$~\ergs, the Eddington limit of a $20M_\sun$ black hole, roughly the most massive compact object that can be formed via core collapse of a single star with near solar metallicity \citep{bel10}. Growing evidence from observations of emission line nebulae around some ULXs indicates that their luminosity is truly high and the emission is not strongly beamed \citep{pak03,kaa04,kaa09}. This requires accretion onto intermediate mass black holes to explain the high luminosity if we assume sub-Eddington radiation.  However, the Eddington limit could be violated under certain circumstances \citep{wat01,beg06} and mild beaming by a factor of a few is expected at high accretion rates \citep{kin01}. These two factors simultaneously could explain the majority of ULXs up to $10^{41}$~\ergs\ as stellar mass black holes \citep{pou07}. The nature of ULXs is still uncertain.

NGC 1313 X-1 was among the first ULXs discovered with Einstein \citep{fab87}. The source has shown significant spectral variability revealed by XMM-Newton observations. Using 12 XMM-Newton observations over a span of a few years, \cite{fen06} reported a tight correlation between the X-ray luminosity and the power-law photon index of the source and a transition from the correlated phase to a soft state. This behavior is similar to that seen in some Galactic black hole X-ray binaries as well as ULXs like NGC 5204 X-1 and Holmberg II X-1 \citep{fen09}, suggesting similar accretion physics, but contrary to some other ULXs like NGC 1313 X-2. \cite{dew10} demonstrated that NGC 1313 X-1 showed distinct temporal properties in different spectral states, manifesting a noisy high state and a quiet low state, interpreted as due to different accretion geometries. A distance of 4.1~Mpc for NGC 1313 is adopted in this paper \citep{men02}.

Optical observations of ULXs offer additional information regarding their nature, and could shed light onto the evolutionary history of the binary, disk geometry, and mode of the mass transfer \citep[e.g.][]{mad08,fen08,kaa09}. Moreover, a conclusion to the stellar vs.\ intermediate mass debate for ULXs would come from dynamical mass measurements, which require identification of optical counterparts.  NGC 1313 X-1 is an important member of the ULX population, and has been extensively studied in X-rays.  However, no optical counterpart of the source has been identified so far.  Here, using data in the archive of the Hubble Space Telescope (HST) and Chandra X-ray Observatory, we report the identification of the optical counterpart to NGC 1313 X-1 and discuss the physical implications.

\section{Astrometry and the Optical Counterparts}

NGC 1313 X-1 has been observed frequently with HST using the Advanced Camera for Surveys (ACS) and with Chandra using the Advanced CCD Imaging Spectrometer (ACIS). A log of HST observations used in the paper is shown in Table~\ref{tab:hst}.

%%%%%%%%%%%%%%%%%%%%%%%%%%%%%%%%%%%%%%%%%%%%%%%%%%%%%%%%%%%%%%
\begin{deluxetable}{lllcl}
\tablecolumns{4} 
\tablewidth{0pc}
\tablecaption{HST/ACS Observations Used in the Paper
\label{tab:hst}}
\tablehead{
\colhead{Date} & \colhead{Start Time} & \colhead{Dateset} & \colhead{Exposure (s)} & \colhead{Filter}  }
\startdata
2003 Nov 17 & 05:52:14\tablenotemark{a} &J8OL01010&1160&F814W \\
				& 06:04:50\tablenotemark{a} &J8OL01030\tablenotemark{b}&1160&F555W \\
 				& 07:30:30 &J8OL01040&2520&F435W \\
				& 09:17:54 &J8OLA1010\tablenotemark{c}&2760&F330W \\
\noalign{\smallskip}\hline\noalign{\smallskip}
2004 Feb 22 & 02:38:19 &J8OL05010&2400&F555W \\
\noalign{\smallskip}\hline\noalign{\smallskip}
2004 Jul 17 & 00:50:56 &J8PH05010&680 &F435W \\
				& 01:08:16 &J8PH05020&680 &F555W \\
				& 01:42:19 &J8PH05030&676 &F814W \\
\noalign{\smallskip}\hline\noalign{\smallskip}
2004 Oct 30 & 06:57:00 &J8YY06011&1381&F814W \\
				& 07:25:29 &J8YY06021&1062&F606W \\
\noalign{\smallskip}\hline\noalign{\smallskip}
2004 Dec 18 & 21:11:19 &J8PH07020\tablenotemark{b}&680 &F555W
\enddata
\tablenotetext{a}{Sub-exposures with different filters were performed alternately.}
\tablenotetext{b}{Observations used for astrometry correction; J8OL01030 is also used for photometry.}
\tablenotetext{c}{Observation with the aperture of the High Resolution Camera, while others are with the Wide Field Camera.}
\end{deluxetable}
%%%%%%%%%%%%%%%%%%%%%%%%%%%%%%%%%%%%%%%%%%%%%%%%%%%%%%%%%%%%%%

\begin{figure}
\centering
\includegraphics[width=0.7\columnwidth]{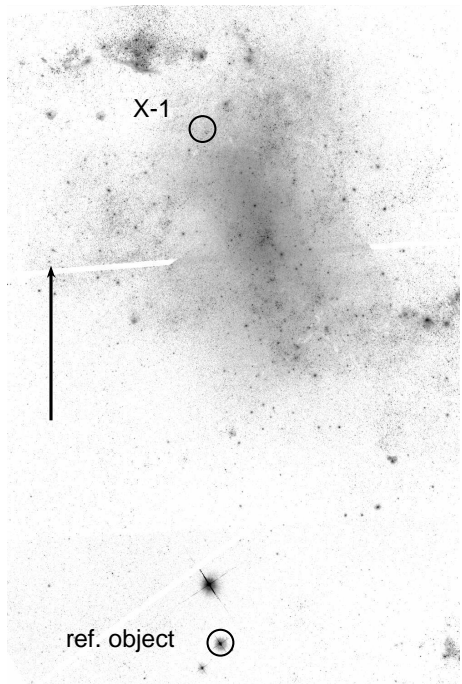}
\caption{The mosaic HST image in F555W containing X-1 and the reference object used for astrometry correction. The arrow points north and has a length of 1\arcmin.
\label{fig:mosaic}}
\end{figure}

\begin{figure}
\centering
\includegraphics[width=0.6\columnwidth]{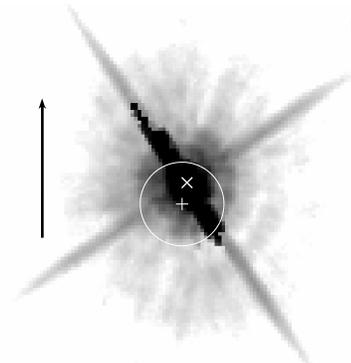}
\caption{Optical image of the reference object. The plus and the circle indicate the Chandra position and error circle of 0.6\arcsec\ radius, respectively. The cross indicates the optical center of the source. The arrow points north and has a length of 1\arcsec.
\label{fig:ref}}
\end{figure}

Chandra observation 2950, with an exposure of 20~ks, is the deepest in this field and thus chosen for astrometry correction. The ULX is located on chip S3 of ACIS with a moderate offset to the optical axis of 2.4\arcmin. We performed source detection using the {\tt wavdetect} tool in CIAO 4.1.2 on a flux-corrected image created using an exposure map assuming a power-law spectrum with a photon index of 1.7 modified by Galactic absorption with $N_{\rm H} = 3.97 \times 10^{20}$~cm$^{-2}$. We then searched in HST images for possible optical counterparts to all X-ray sources detected on the S3 chip, and found one good candidate with both X-ray and optical emission, hereafter referred to as the reference object (see Figure~\ref{fig:mosaic} \& \ref{fig:ref}).

The relative astrometry between Chandra and HST can be improved using the reference object. Due to the relatively small field of view of HST, the reference object and X-1 do not appear on any single ACS image. We thus created a mosaic image from two ACS observations with the Wide Field Camera with the F555W filter, datasets J8PH01030 and J8PH07020. These two observations have an overlap of about 6 arcmin$^2$. Stars in the overlap region were detected using the {\tt daofind} task, cross identified using {\tt xyxymatch}, then used to calculate a relative shift and rotation between the two images using {\tt geomap}.  Figure~\ref{fig:mosaic} shows a mosaic image created using the {\tt multidrizzle} task with these parameters.  The mosaic image was then aligned to stars in the Two Micron All Sky Survey \citep[2MASS;][]{skr06} using the Graphical Astronomy and Image Analysis Tool \citep[GAIA;][]{dra09}. We find 11 2MASS counterparts that appear point-like, isolated, and bright ($K \leq 15$) on the mosaic image. Fitting to them results in a root-mean-square average deviation less than 2 pixels ($\approx 0.1\arcsec$), and an absolute astrometric uncertainty of 0.07\arcsec\ at 90\% confidence \citep[also see][]{fen08}.  We note that this uncertainty does not influence the alignment of the X-ray and optical images.

%%%%%%%%%%%%%%%%%%%%%%%%%%%%%%%%%%%%%%%%%%%%%%%%%%%%%%%%%%%%%%
%%%%%%%%%%%%%%%%%%%%%%%%%%%%%%%%%%%%%%%%%%%%%%%%%%%%%%%%%%%%%%

%%%%%%%%%%%%%%%%%%%%%%%%%%%%%%%%%%%%%%%%%%%%%%%%%%%%%%%%%%%%%%
%%%%%%%%%%%%%%%%%%%%%%%%%%%%%%%%%%%%%%%%%%%%%%%%%%%%%%%%%%%%%%

%%%%%%%%%%%%%%%%%%%%%%%%%%%%%%%%%%%%%%%%%%%%%%%%%%%%%%%%%%%%%%
\begin{figure*}
\centering
\includegraphics[width=0.32\textwidth]{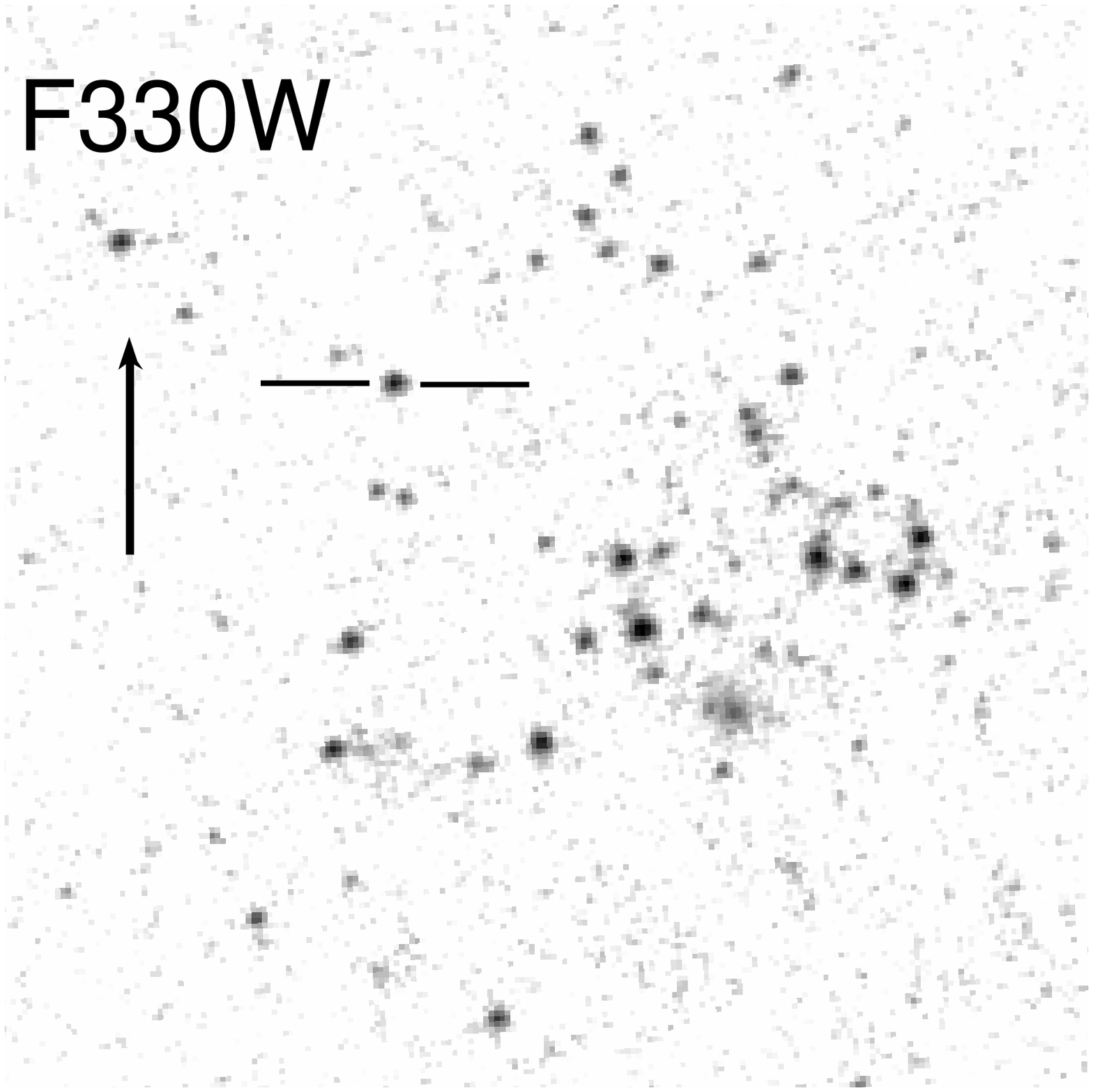}
\includegraphics[width=0.32\textwidth]{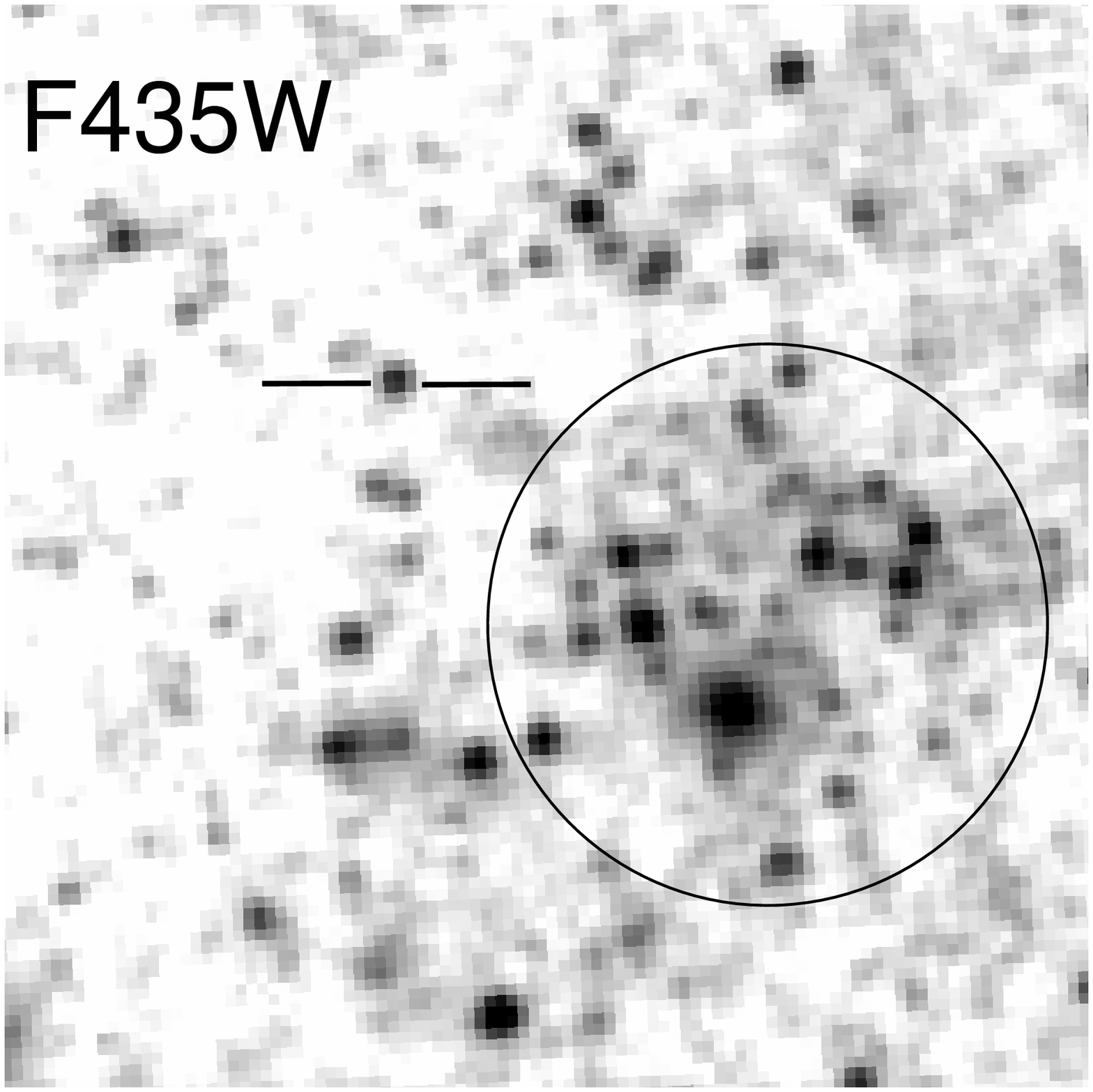}
\includegraphics[width=0.32\textwidth]{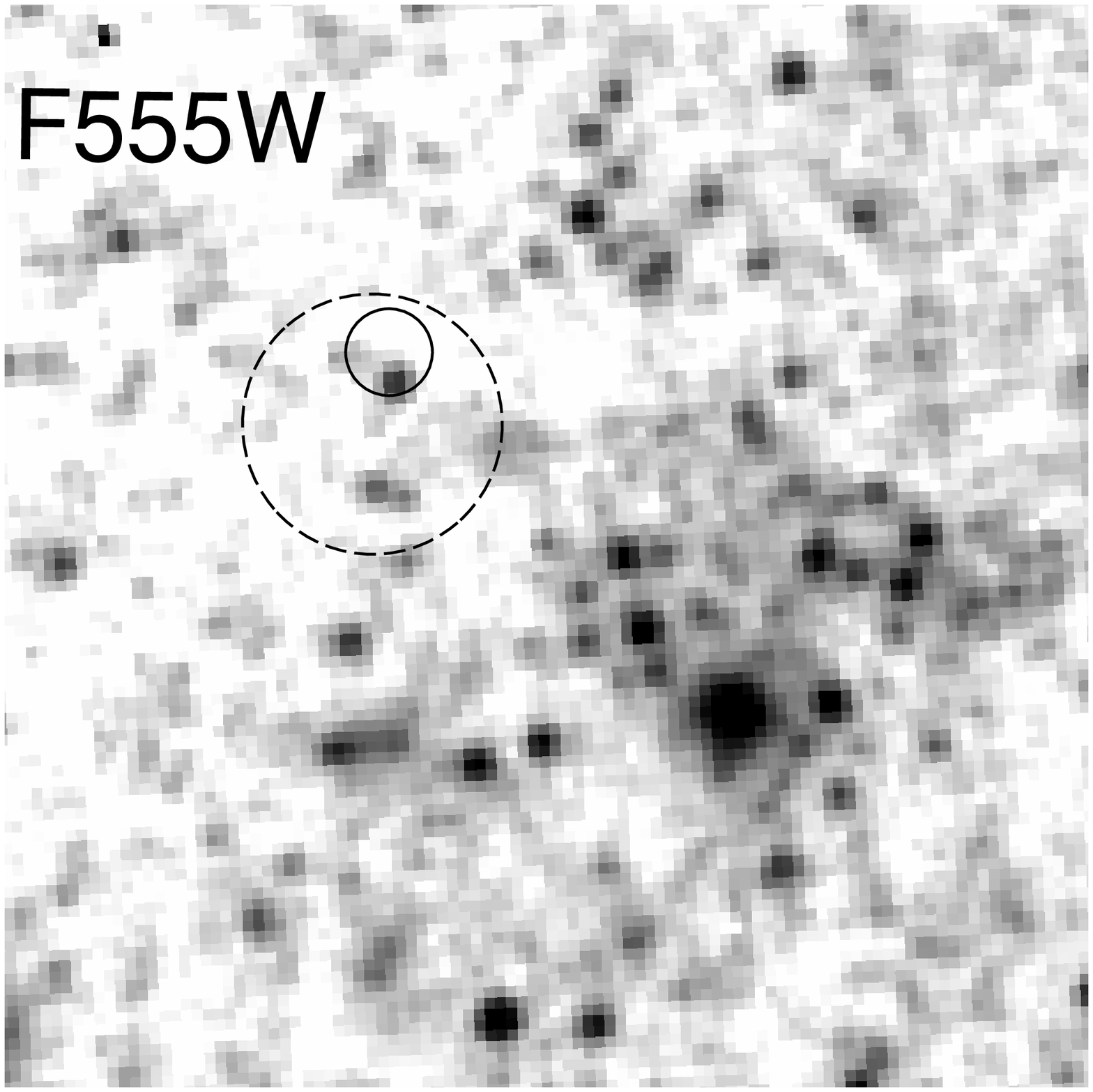}\\
\includegraphics[width=0.32\textwidth]{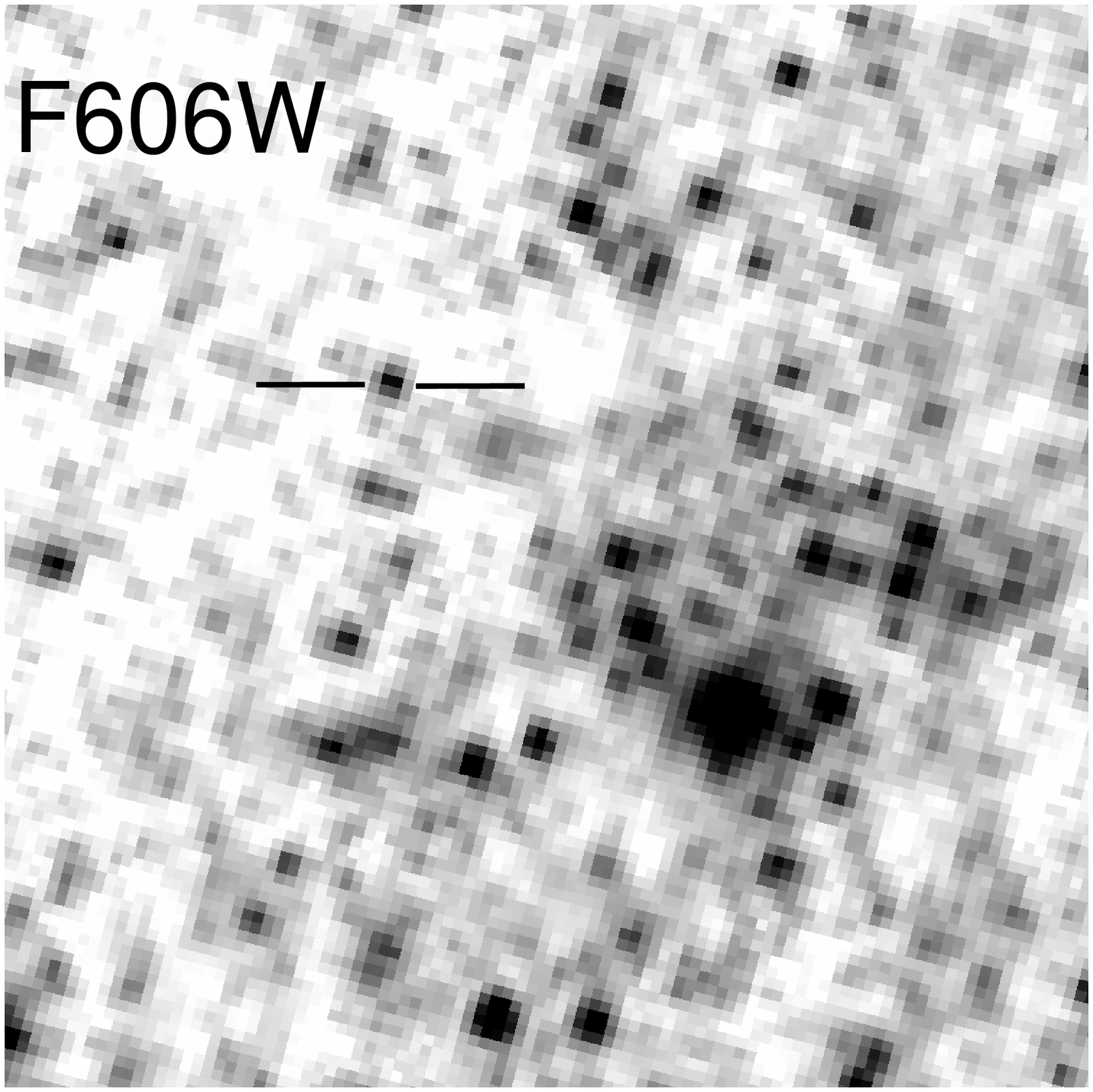}
\includegraphics[width=0.32\textwidth]{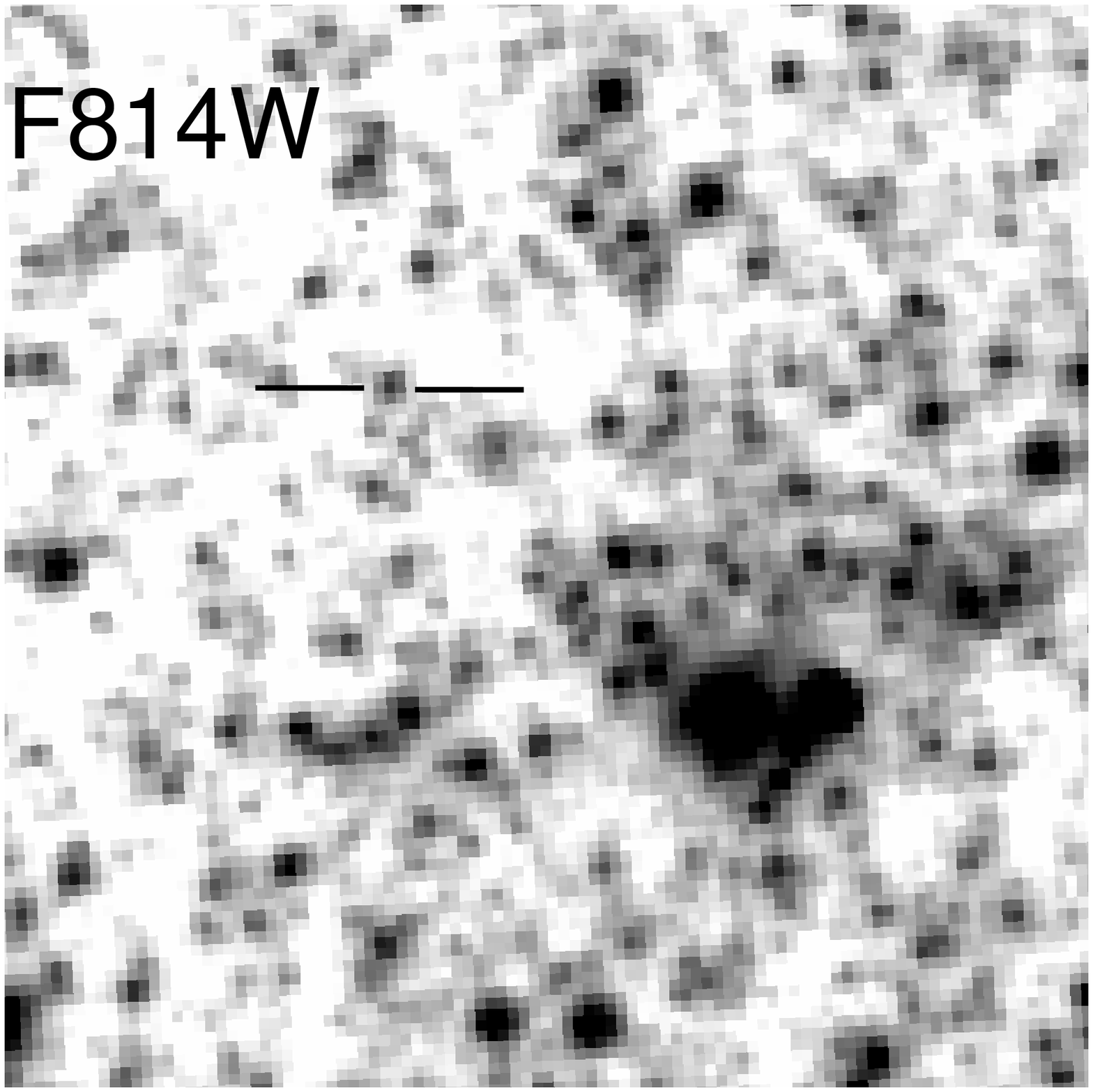}
\caption{HST images around NGC 1313 X-1 in different bands. On the F555W image the dashed and solid circles indicate, respectively, the original and corrected Chandra position errors of X-1. On F435W, the circle indicates a group of stars nearby. The horizontal bars point to the bright optical object in the corrected X-ray error circle. The F606W image is from the observation on 2004 Oct 30, and others are obtained from observations on 2003 Nov 17.  All arrows point north and have a length of 1\arcsec.
\label{fig:x1}}
\end{figure*}
%%%%%%%%%%%%%%%%%%%%%%%%%%%%%%%%%%%%%%%%%%%%%%%%%%%%%%%%%%%%%%

The reference object appears as a saturated star on the HST image. Its optical position is estimated from the cross point of the diffraction spikes as $\rm R.A. = 03^h18^m18\fs858$ $\rm decl. = -66\arcdeg32\arcmin29\farcs92$ (J2000.0) with a negligible statistic error of about 0.01\arcsec.  It has 82 photons detected on the Chandra image with a significance of 38$\sigma$ and an error radius of 0.17\arcsec\ at 90\% confidence.  Using MARX simulation, the systematic offset in the position calculated by {\tt wavdetect} caused by asymmetry of the photon distribution is estimated as $\Delta$R.A.\ = 0.03\arcsec\ and $\Delta$decl.\ = $-0.04$\arcsec.  These values were used to correct the offset in the coordinates from {\tt wavdetect}.  In Figure~\ref{fig:ref}, the optical image of the reference object is shown.  The cross indicates the optical center and the plus indicates the X-ray position with the circle showing the 0.6\arcsec\ radius absolute astrometry uncertainty of Chandra.  The X-ray image can be registered onto the optical image by aligning the plus and the cross.

For X-1, there are 2968 photons detected with Chandra, and the statistic error radius is 0.04\arcsec\ at 90\%.  The systematic errors are $\Delta$R.A.\ = $-0.08$\arcsec\ and $\Delta$decl.\ = 0.02\arcsec\ and are taken into account.  The corrected position of X-1 relative to HST is $\rm R.A. = 03^h18^m20\fs013$ $\rm decl. = -66\arcdeg29\arcmin10\farcs79$ (J2000.0). The uncertainty on this corrected position is similar to the statistical error on the X-ray location of the reference object, i.e.\ 0.17\arcsec, since all other errors are significantly smaller. However, the rotation of the X-ray image cannot be fixed with a single object, and the accuracy in rotation of Chandra astrometry is not discussed in any available Chandra documentation. We estimated the rotation accuracy from a few Chandra observations with multiple optically identified X-ray sources and obtained a typical rotation about 2\arcmin\ between Chandra and 2MASS. Taking this value into account, with a distance of $\sim$200\arcsec\ between the reference object and X-1, the error caused by rotation is about 0.1\arcsec. Combining the two errors, 0.1\arcsec\ and 0.17\arcsec, in quadrature, we estimate that the corrected X-ray position has an uncertainty of about 0.2\arcsec\ at the 90\% confidence level. 

The F555W image around X-1 is shown in Figure~\ref{fig:x1}, on which a large, dashed circle indicates the original Chandra position with an absolute error of 0.6\arcsec, while a small, solid circle indicates the corrected relative position with a radius of 0.2\arcsec.  The astrometry correction successfully rules out a few objects as possible counterparts to X-1, but leaves one relatively bright object within the corrected relative error circle and another faint feature just to the east of the circle.  Even if the error radius of the corrected Chandra position is 2.5 times larger, i.e.\ 0.5\arcsec, this conclusion does not change. All available HST images of X-1 in other bands (F330W, F435W, F606W, and F814W) are also displayed with the same scale as the F555W image. Horizontal bars are plotted pointing to the bright object in the error circle. The bright object is obvious in all bands. The faint feature is more significant at shorter wavelength. 

At about 2\arcsec\ to the southwest of X-1, a group of stars with an extent of about 1.5\arcsec\ (30 pc) are obviously seen, indicated by a circle on the F435W image. A few bright objects in the group may consist of multiple unresolved sources, especially on images of long wavelength.

\section{Photometry}

\subsection{Flux and Variability of the counterpart}

The region around X-1 on the HST images is crowded, so we decided to perform aperture photometry with an aperture radius of 0.1\arcsec, in order to reduce contamination from nearby sources. To test the aperture correction for this relatively small aperture, we selected nearby isolated bright objects and performed photometry using both small and large apertures.  We found a typical uncertainty around 0.02~mag, which is added into the uncertainty in our reported magnitudes.  We note that PSF photometry in this circumstance does not offer more reliable measurement, as features near the source have somewhat complex morphology and are not point-like.  Background levels were estimated from a concentric annulus where no sources appear; if a couple dim sources could not be avoided, we then chose a background region as large as possible to depress the source contribution.  The intrinsic source flux or magnitude was calculated using the SYNPHOT package using the count rate obtained from photometry corrected to an infinite aperture \citep{sir05} and Galactic reddening $E(B-V) = 0.11$ \citep{sch98}.  The PSF may vary slightly with temperature\footnote{http://www.stsci.edu/hst/acs/documents/isrs/isr0712.pdf}, causing a systematic shift about 0.02-0.05~mag between observations with such a small aperture.  Several nearby isolated bright stars were used to calculate a zero point shift to correct for this effect using the magnitudes measured on 2003 Nov 17 as reference.  The extinction corrected fluxes at different bands were fitted to a power-law spectrum, $F_\nu \propto \nu^\alpha$, where $\nu$ is the pivot frequency of the filter in Hz and $F_\nu$ is the flux in erg~s$^{-1}$~cm$^{-2}$~Hz$^{-1}$.  The same power-law spectrum was assumed during the conversion from rate to flux; a couple cycles of iteration were needed here to produce a consistent power-law index before and after the conversion.

%%%%%%%%%%%%%%%%%%%%%%%%%%%%%%%%%%%%%%%%%%%%%%%%%%%%%%%%%%%%%%
\begin{deluxetable}{lllll}
\tablecolumns{5} 
\tablewidth{0pc}
\tablecaption{Magnitudes corrected for Galactic extinction of the bright optical object in the corrected error circle of X-1.
\label{tab:mag}}
\tablehead{
\colhead{Band} & \colhead{2003 Nov 17} & \colhead{2004 Feb 22} & \colhead{2004 Jul 17} & \colhead{2004 Oct 30}}
\startdata
\multicolumn{5}{c}{ST magnitudes in HST bands} \\
\noalign{\smallskip}\hline\noalign{\smallskip}
F330W & $22.26\pm0.03$ & \nodata & \nodata & \nodata \\
F435W & $23.00\pm0.03$ & \nodata & $23.11\pm0.04$ & \nodata \\
F555W & $23.65\pm0.04$ & $23.68\pm0.04$ & $23.90\pm0.05$ & \nodata \\
F606W & \nodata & \nodata & \nodata & $24.08\pm0.04$ \\
F814W & $24.92\pm0.06$ & \nodata & $24.86\pm0.06$ & $24.92\pm0.07$ \\
\noalign{\smallskip}\hline\noalign{\smallskip}
\multicolumn{5}{c}{Vega magnitudes in Johnson-Cousins bands} \\
\noalign{\smallskip}\hline\noalign{\smallskip}
$U$  	&$22.62\pm0.03$	 &\nodata			 &\nodata			 &\nodata				 \\
$B$  	&$23.64\pm0.03$	 &\nodata			 &$23.74\pm0.04$   &\nodata				 \\
$V$  	&$23.72\pm0.04$	 &$23.74\pm0.04$   &$23.96\pm0.05$   &$23.85\pm0.04$  	 \\
$I$  	&$23.65\pm0.06$	 &\nodata			 &$23.59\pm0.06$   &$23.65\pm0.07$  	 
\enddata
\end{deluxetable}
%%%%%%%%%%%%%%%%%%%%%%%%%%%%%%%%%%%%%%%%%%%%%%%%%%%%%%%%%%%%%%

\begin{figure}
\centering
\includegraphics[width=\columnwidth]{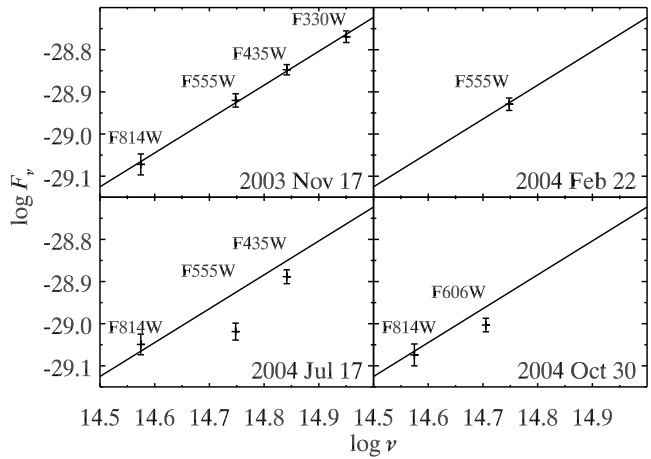}
\caption{Multiband spectrum of the bright optical object in the corrected error circle of X-1 from different epochs. The lines in different panels are the same, indicating the best-fit power-law model, $F_\nu \propto \nu^{0.81 \pm 0.07}$ to the data on 2003 Nov 17.
\label{fig:spec}}
\end{figure}

\begin{figure}
\centering
\includegraphics[width=0.65\columnwidth]{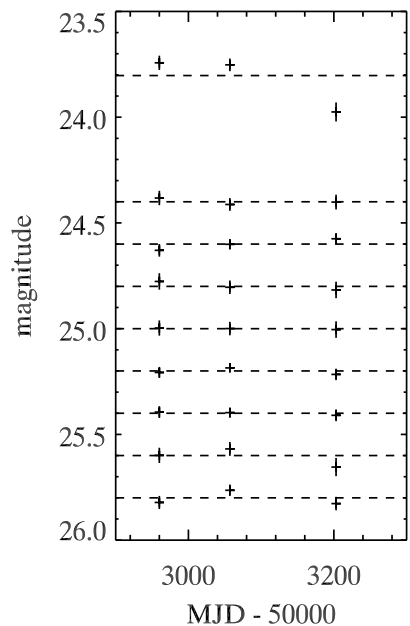}
\caption{Light curves in the F555W band for the bright optical object in the corrected error circle of X-1 and another 8 nearby sources within 10\arcsec\ with similar magnitude. The instrumental magnitude is computed as $m = 25 - 2.5\log r$, where $r$ is the observed net count rate. The top one is the counterpart of X-1, and the magnitudes for other sources are shifted vertically for clarity. The dashed lines are the best fit constants for each.
\label{fig:var}}
\end{figure}

ST magnitudes corrected for Galactic extinction for the bright object are listed in Table~\ref{tab:mag}. They are also transformed into Vega magnitudes in the closest Johnson $UBV$ and Cousins $I$ bands \citep{sir05}. Additional errors introduced by the uncertainty on the assumed power-law index is taken into account but are not significant. The source has an absolute visual magnitude $M_V = -4.35 \pm 0.04$ and color $(B-V)_0 = -0.08 \pm 0.05$ (Johnson magnitudes with VEGA zero points corrected for Galactic extinction) from observations on 2003 Nov 17.  The intrinsic fluxes in different bands from different epochs are plotted in Figure~\ref{fig:spec}, where the solid lines indicate the best power-law model fitted to the spectrum on 2003 Nov 17.  As shown in Figure~\ref{fig:spec}, the fluxes measured on 2004 Jul 17 and Oct 30 show a deviation from the best-fit power-law obtained from 2003 Nov 17, except in the F814W band. The largest variation is seen in the F555W filter with a magnitude increase, $\Delta m = 0.25 \pm 0.06$ (here the aperture correction error is not included as we are considering relative shifts for the same filter), for the observation on 2004 Jul 17 with respect to that on 2003 Nov 17. Fitting to a constant with the three F555W magnitudes results in $\chi^2 = 23.4$ for 2 degrees of freedom (d.o.f.).  To investigate whether the variation is intrinsic or instrumental, we selected 8 sources with similar magnitude in F555W within 10\arcsec\ of X-1, and checked their magnitude variation (again, without considering the aperture correction error), shown in Figure~\ref{fig:var}. All these 8 sources show a light curve consistent with a constant, with $\chi^2 = 0.3 - 3.3$ for 2 d.o.f. Also, the multiband spectra of these 8 sources are closely consistent between the two observations on 2003 Nov 17 and 2004 Jul 17. Therefore, we conclude that the variation in the F555W band for the counterpart is intrinsic. We also checked the variability in F435W and F814W for the ULX and comparison sources in the same way. Within the photometric errors, there is no evidence for variability of the ULX. However, the data in the F435W band are incompatible with the level of variability seen in the F555W band only at the 90\% confidence level.

%%%%%%%%%%%%%%%%%%%%%%%%%%%%%%%%%%%%%%%%%%%%%%%%%%%%%%%%%%%%%%
%%%%%%%%%%%%%%%%%%%%%%%%%%%%%%%%%%%%%%%%%%%%%%%%%%%%%%%%%%%%%%

%%%%%%%%%%%%%%%%%%%%%%%%%%%%%%%%%%%%%%%%%%%%%%%%%%%%%%%%%%%%%%
%%%%%%%%%%%%%%%%%%%%%%%%%%%%%%%%%%%%%%%%%%%%%%%%%%%%%%%%%%%%%%

For the faint object to the east of the corrected error circle, it has a reddening corrected ST magnitude of $25.3 \pm 0.1$ in F555W on 2003 Nov 17. Interestingly, there seems to be an extended feature connecting it to the bright object, as one can see at long wavelengths, in particular in F814W, while the object itself, at the tip of this extended feature, becomes invisible in F814W. This faint object shows a constant flux within errors. Due to the deepness of HST observations, such a faint object has a relatively high chance probability to show up in any circle of 0.2\arcsec-radius. We currently cannot constrain its nature or rule out the possibility that it is not the physical counterpart to X-1. However, due to two facts we prefer the bright object as the better candidate of the counterpart to X-1. First, its magnitude and spectral shape are similar to the optical counterparts of most other ULXs that have a unique identification in optical \citep{tao11}. Also, variability is expected for an accreting source.

\subsection{X-ray to Optical Flux Ratio}

Two sets of HST observations used here have simultaneous Chandra observations.  Unfortunately, X-1 caused severe pileup on the CCD with about 25\% of events being piled from multiple photons in these observations, hampering us from obtaining reliable spectra and fluxes. Based on the recorded X-ray count rates and using PIMMS, we find that the X-ray flux of the source during the two Chandra observations is within the range found by XMM-Newton \citep{fen06}. Therefore, we adopt this range as the X-ray flux to calculate the X-ray to optical flux ratio, defined as $\log(f_{\rm X}/f_V) = \log f_{\rm X} + m_V/2.5 + 5.37$, where $f_{\rm X}$ is the observed X-ray flux in the 0.3-3.5 keV band and $m_V$ is the observed visual magnitude \citep{mac88}. Inserting $f_{\rm X} = (1.4 - 4.6) \times 10^{-12}$~\ergcms, found from the flux range in \citet{fen06}, and $m_V = 24.1$, we have $\log(f_{\rm X}/f_V) = 3.1-3.7$. This is significantly larger than the X-ray to optical flux ratio of active galactic nuclei (AGN), which is usually found in the range from $-1$ to 1.7 \citep{mac88,sto91}.

For the purpose of distinguishing low mass and high mass X-ray binaries, an alternative definition of the ratio, called X-ray to optical color index, has been proposed as $\xi = B_0 + 2.5\log F_{\rm X}$, where $B_0$ is the extinction corrected $B$ magnitude and $F_{\rm X}$ is the observed X-ray flux density in 2-10 keV in units of $\mu$Jy \citep{par81,par95}.  With $B_0 = 23.6$ and $F_{\rm X} = 0.03-0.23$~$\mu$Jy \citep[also derived from][]{fen06}, $\xi$ is estimated to be in the range of 20-22 for NGC 1313 X-1. For high mass X-ray binaries during outbursts, $\xi$ is around 12-18; for low mass X-ray binaries (LMXBs), the $\xi$ distribution peaks around 21-22, with a deviation of 1-2 \citep{par95}. Therefore, in terms of $\xi$, the ULX is more similar to LMXBs.

\subsection{Color-Magnitude Diagram}

\begin{figure}
\centering
\includegraphics[width=\columnwidth]{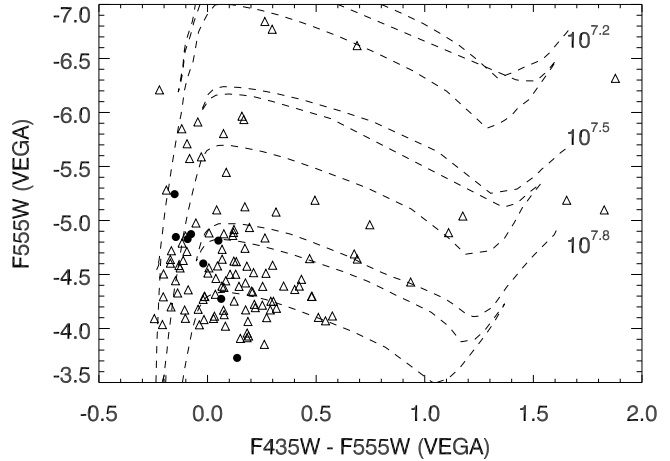}
\caption{Color-magnitude diagram for objects within 10\arcsec\ (200~pc) of NGC 1313 X-1. Dashed lines are theoretical isochrones at ages of $10^{7.2}$, $10^{7.5}$, and $10^{7.8}$~yr, respectively. Filled circles are for objects in the nearby group of stars, i.e.\ inside the circle on F435W in Figure~\ref{fig:x1}.
\label{fig:cmd}}
\end{figure}

The magnitudes for objects within 10\arcsec\ of X-1 (about 200~pc correspondingly) were calculated using the DAOPHOT package in the Interactive Data Language (IDL). A total number of 123 bright sources with a flux 4$\sigma$ above the background were detected using {\tt daofind} from observations on 2003 Nov 17 in the F435W and F555W bands. All these sources have a local fit $\chi^2 < 3$ to exclude extended sources and image defects. We performed PSF photometry for these sources and created a color-magnitude diagram shown in Figure~\ref{fig:cmd}, for F555W vs.\ F435W$-$F555W. The magnitudes are computed assuming an A0 spectrum based on their mean color, normalized to VEGA zero point, and corrected for the Galactic extinction. Typical errors are 0.03 for magnitudes and 0.05 for colors. For the nearby group of stars, eight bright, point-like sources are detected and displayed in the figure as filled circles.

%%%%%%%%%%%%%%%%%%%%%%%%%%%%%%%%%%%%%%%%%%%%%%%%%%%%%%%%%%%%%%
%%%%%%%%%%%%%%%%%%%%%%%%%%%%%%%%%%%%%%%%%%%%%%%%%%%%%%%%%%%%%%

Observations of H~{\sc ii} regions in NGC 1313 suggest a flat distribution of oxygen abundance with $12 + \log({\rm O}/{\rm H}) = 8.4$ across the disk \citep{wal97}, converting to a metallicity of $Z = 0.006$ based on standard solar composition \citep{gre98}. Theoretical isochrones\footnote{Obtained from http://stev.oapd.inaf.it/cmd} of stars with the same metallicity at ages of $10^{7.2}$, $10^{7.5}$, and $10^{7.8}$~yr are plotted \citep{mar08}. As one can see, most objects are older than $10^{7.5}$~yr, and almost none of them is younger than $10^{7.2}$~yr. In the diagram, stars in the nearby group do not appear differently from field sources. We also checked the diagram between F555W and F814W and obtained the same conclusion.

\section{Discussion}

We have demonstrated the identification of the optical counterpart of NGC 1313 X-1 as the brighter object in the corrected error circle. Deviation from a single power-law spectrum on 2004 Jul 17 may suggest that at least two components are responsible for the optical emission. Due to a constant flux in F814W and variability in F555W, the red wing of the spectrum is possibly dominated by emission from the companion star and the blue wing is likely dominated by disk radiation.

Assuming that the red component is mainly due to star light from the companion, we can place constraint on its absolute magnitudes $M_V \gtrsim -4.1$, $M_I =-4.5 $, and color $(V-I)_0 \gtrsim 0.4$, where the values are translated from F555W and F814W on 2004 Jul 17 and the inequality accounts for the fact that the F555W flux of the red component is lower than the observed value. A K5-M0 II bright giant would match the $I$-band magnitude and be consistent with the color. The F814W flux on 2003 Nov 17 should also be due to the companion emission under this hypothesis, and the single power-law is then a composition of two components. The distinct optical spectra between 2003 Nov 17 and 2004 Jul 17 indicate that the disk emission has a varying break frequency $\nu_{\rm b} \approx (5-8) \times 10^{14}$~Hz, corresponding to a break temperature $T_{\rm b} = h\nu_{\rm b}/k \approx 3 \times 10^4$~K, where $h$ and $k$ are the Planck and Boltzmann constants, respectively.

Disk emission in the optical can be decomposed into two origins, one from intrinsic viscous release of the multicolor disk, and the other from X-ray reprocessing (irradiation) in the outer regions of the disk.  The intrinsic emission of a standard thin disk has a spectral form of $\nu^{1/3}$ at frequencies above $kT_{\rm out}/h$, where $T_{\rm out}$ is the disk temperature at the outermost radius, and then breaks to $\nu^{2}$ at much lower frequencies. The two point power-law index between the F555W and F435W fluxes on 2004 Jul 17 is $\alpha = 1.4 \pm 0.3$ corrected for Galactic extinction; if additional extinction exists in NGC 1313, $\alpha$ would be even larger.  Thus, the blue wing of the spectrum is steeper than $\nu^{1/3}$ and cannot be explained by intrinsic disk emission only, unless the disk is truncated with $T_{\rm out} \approx T_{\rm b}$. For viscous dissipation on a standard thin disk, we have $R_{\rm out}/R_{\rm in} = (T_{\rm out}/T_{\rm in})^{-4/3}$. If we assume $T_{\rm in} = 0.1-1$~keV ($10^6-10^7$~K) for either intermediate or stellar mass black holes, then we get $R_{\rm out}/R_{\rm in} = 100 - 2000$, suggesting the presence of a relatively narrow disk, as has been proposed for NGC 6946 X-1 \citep{kaa10} and SS~433 \citep{per10}. Otherwise, if disk irradiation is dominant, it is expected to have an effective temperature $T_{\rm ir} \approx T_{\rm b}$ with a surface area around $10^{24}$~cm$^2$ to reach the observed flux density. Therefore, a disk size about $10^{12}$~cm is inferred. 

As optical fluxes at different bands were not obtained simultaneously, an alternative explanation of the non-power-law spectrum seen on 2004 Jul 17 could be due to short-term variation of the disk emission. The last sub-exposure of F435W started in advance of the F555W observation by 544~s, which, corresponding to a light cross distance over $10^{13}$~cm, is sufficiently long to respond to the rapid X-ray variation. Strong, fast optical variations have been found for both LMXBs and ULXs \citep{bra83,gri08}, and NGC 1313 X-1 is highly variable in X-rays at short timescales \citep{dew10}. Simultaneous X-ray and optical observations in the future could test different scenarios.

The properties of the companion star could be constrained by comparing the observed $M_V$ and $(B-V)_0$ with theoretical values derived from binary evolution \citep{pat08,pat10}. The data can be best-fitted with a donor of 10-15~$M_\sun$ during hydrogen shell burning given a black hole mass of 10-100~$M_\sun$, or during main sequence only if the black hole mass is close to 100~$M_\sun$. Also, it can be ruled out that the donor starts to fill the Roche-lobe during the giant phase if the black hole mass is greater than 20~$M_\sun$ \citep{pat10}.

The color-magnitude diagram and isochrones reveal a stellar population around NGC 1313 X-1 with most stellar ages greater than $10^{7.5}$~Myr.  The nearby region (within 10\arcsec\ or 200~pc) also lacks very blue stars and young star clusters.  This places a constraint on the mass of the companion star as no more than $\sim 10 M_\sun$.  This is consistent with the hypothesis above that the companion may be a late type bright giant. The ULX is close to a group of stars, which has an extent of about 30 pc and has a similar age to field stars. Whether or not the ULX belongs to the star group is unknown. This is different from the environment around NGC 1313 X-2, which is found to be associated with a young group consisting of bright and blue stars \citep{gri08}.

\acknowledgments We thank the referee for useful comments that have improved the paper. %HF acknowledges funding support from the National Natural Science Foundation of China under grants 10903004 and 10978001, the 973 Program of China under grant 2009CB824800, the Foundation for the Author of National Excellent Doctoral Dissertation of China under grant 200935, the Tsinghua University Initiative Scientific Research Program, and the Program for New Century Excellent Talents in University. PK acknowledges financial support from NASA grant NNX08AJ26G.

{\it Facility:} \facility{CXO(ACIS)}, \facility{HST(ACS)}

\end{document}